\documentclass[reprint, twocolumn, 
superscriptaddress, 
amsmath, amssymb, aps, 
]{revtex4}
\usepackage{latexsym}    
\usepackage{graphicx}   
\usepackage{verbatim}   
\usepackage[normalem]{ulem} 
\usepackage{xcolor}     
\usepackage{tikz}       
\usetikzlibrary{shapes}
\usepackage{subfigure}  
\usepackage{url}
\usepackage{hyperref}   
\usepackage{tensor}     
\usepackage{amssymb}
\usepackage{enumerate}	
\usepackage{braket}		
\usepackage{physics}
\usepackage{color}

\newcommand{\DBBrown}{\color [rgb]{ .52, .30, .19}}

\newcommand{\DBLGrey}{\color [rgb]{ .52, .58, .63}}

\newcommand{\DHY}[1]{{\DBBrown #1}}                

\newcommand{\COMDHY}[1]{{\DBLGrey \sout{#1}}}  

\newcommand{\PCOLDAGREE}[1]{{#1}}                  
\newcommand{\PCOLD}[1]{{#1}}                       

\newcommand{\COMPCOLDAGREE}[1]{{}}                 
\newcommand{\COMPCOLD}[1]{{}}                      

\newcommand{\HWCOLD}[1]{{#1}}                      
\newcommand{\HWC}[1]{{#1}}                         
\newcommand{\COMHWCOLD}[1]{{}}                     
\newcommand{\COMHWC}[1]{{}}                        
\newcommand{\PC}[1]{{#1}}                          

\newcommand{\COMPC}[1]{{}}                         

\newcommand{\HWCNEW}[1]{{#1}}                         
\newcommand{\COMHWCNEW}[1]{{}}                        

\begin{document}

\title{\COMPCOLDAGREE{Analysis of generic unitary black-hole evaporation models from first principles}\PCOLDAGREE{A generic unitary black-hole evaporation model based on first principles}}

\author{Kuan-Yu Chen}
\email{r07222014 [at] ntu.edu.tw}
\affiliation{Leung~Center~for~Cosmology~and~Particle~Astrophysics, National~Taiwan~University, 
Taipei~10617, Taiwan, R.O.C.\\}
\affiliation{Department of Physics and Center for Theoretical \COMHWCOLD{Sciences}\HWCOLD{Physics}, National~Taiwan~University, 
Taipei~10617, Taiwan, R.O.C.\\}

\author{Pisin Chen}
\email{pisinchen [at] phys.ntu.edu.tw}
\affiliation{Leung~Center~for~Cosmology~and~Particle~Astrophysics, National~Taiwan~University, 
Taipei~10617, Taiwan, R.O.C.\\}
\affiliation{Department of Physics and Center for Theoretical \COMHWCOLD{Sciences}\HWCOLD{Physics}, National~Taiwan~University, 
Taipei~10617, Taiwan, R.O.C.\\}
\affiliation{Graduate~Institute~of~Astrophysics, National~Taiwan~University, 
Taipei~10617, Taiwan, R.O.C.\\}
\affiliation{Kavli~Institute~for~Particle~Astrophysics~and~Cosmology, SLAC~National~Accelerator~Laboratory, Stanford~University, 
Stanford, CA~94305, U.S.A.\\}

\author{Hsu-Wen Chiang}
\email{b98202036 [at] ntu.edu.tw}
\affiliation{Leung~Center~for~Cosmology~and~Particle~Astrophysics, National~Taiwan~University, 
Taipei~10617, Taiwan, R.O.C.\\}
\affiliation{Department of Physics and Center for Theoretical \COMHWCOLD{Sciences}\HWCOLD{Physics}, National~Taiwan~University, 
Taipei~10617, Taiwan, R.O.C.\\}

\author{Dong-Han~Yeom}
\email{innocent.yeom [at] gmail.com}
\affiliation{Department of Physics Education, Pusan National University, Busan 46241, Republic of Korea\\}
\affiliation{Research Center for Dielectric and Advanced Matter Physics,
Pusan National University, Busan 46241, Republic of Korea\\}

\begin{abstract}
\COMHWC{\COMDHY{This is ``$\backslash$COMDHY\{...\}''.} \DHY{This is ``$\backslash$DHY\{...\}''.} \COMPCOLD{This is ``$\backslash$COMPCOLD\{...\}''.} \PCOLD{This is ``$\backslash$PCOLD\{...\}''.} \COMPCOLDAGREE{This is ``$\backslash$COMPCOLDAGREE\{...\}''.} \PCOLDAGREE{This is ``$\backslash$PCOLDAGREE\{...\}''.} \HWCOLD{This is ``$\backslash$HWCOLD\{...\}''.} \COMHWCOLD{This is ``$\backslash$COMHWCOLD\{...\}''.} \HWC{This is ``$\backslash$HWC\{...\}''.}\COMHWC{This is ``$\backslash$COMHWC\{...\}''.} \HWCAGREE{This is ``$\backslash$HWCAGREE\{...\}''.}  \COMHWCAGREE{This is ``$\backslash$COMHWCAGREE\{...\}''.} \COMPC{This is ``$\backslash$COMPC\{...\}''.} \PC{This is ``$\backslash$PC\{...\}''.}}
\HWC{Based on the discretized horizon picture, w}\COMHWC{W}e introduce a \COMHWC{generic}\COMPC{\HWC{versatile} }macroscopic \HWC{effective} model \HWC{of the horizon area quanta} that encapsulates the features necessary for black holes to evaporate consistently. \HWC{The price to pay is the introduction of a ``hidden sector'' that represents our lack of knowledge about the final destination of the black hole entropy. We focus on the peculiar form of the interaction between this hidden sector and the black hole enforced by the self-consistency.} Despite the expressive power of the model, we arrive at several qualitative statements.
Furthermore, \HWC{we identify} these \COMHWC{macroscopic properties are associated with}\HWC{statements as} features inside the microscopic density of states \HWC{of the horizon quanta}, with \HWC{the dimension of the configuration space being associated with the area per quanta in Planck unit,}\COMHWC{ zero-frequency poles corresponding to the final burst scenario, and} a UV cutoff proportional to the amount of excess entropy relative to Bekenstein's law at the end of evaporation\HWC{, and a zero-frequency-pole-like structure corresponding to, similarly, the amount of excess entropy at IR limit. We then relate this nearly-zero-frequency structure to the soft hairs proposed by Strominger et al., and argue that we should consider deviating away from the zero frequency limit for soft hairs to participate in the black hole evaporation}.
\COMPCOLD{(This abstract is a bit too abstract and too brief. Please revise it so as to make the conclusions
and consequences of our construction more explicit.)}
\end{abstract}

\maketitle

\paragraph*{Introduction.}
As \COMPCOLD{a crowning achievement}\PCOLD{an inevitable consequence} of the semi-classical analysis of black hole (BH), the information loss paradox\COMHWCOLD{ (ILP)} is often considered \cite{Mathur:2009hf,Hawking:1976ra
} a gateway to\COMPCOLD{ward} quantum gravity. The paradox suggests that while derived from quantum field theory in curved spacetime, the Hawking evaporation process \cite{Hawking:1974sw} inevitably leads to an enormous amount of entanglement entropy between \COMHWC{Hawking radiations and \COMPCOLDAGREE{a}\PCOLDAGREE{the} BH with a vanishing mass}\HWC{an evaporated BH and the radiation it released}
, thus violating one or more of the following fundamental assumptions in modern physics: unitarity, locality, and general covariance. Many \PCOLDAGREE{authors} \cite{Susskind:1993if,Wald:1993nt,Parikh:2004ih,Hayden:2007cs,Almheiri:2012rt,Maldacena:2013xja,Strominger:2013jfa,Chen:2014jwq,Hotta:2015yla,Carney:2017jut,Unruh:2017uaw,Almheiri:2019psf,Pasterski:2020xvn} have attempted to solve the paradox 
by various alternations to the classical BH picture. \PCOLDAGREE{The final resolution evidently is \COMHWC{not yet in sight}\HWC{yet to come}.} In this work, instead of providing yet another specific solution, we take the minimalistic approach and aim at \COMPCOLDAGREE{the}\PCOLDAGREE{establishing a} generic mathematical structure \COMHWC{and }\COMPCOLDAGREE{the resulting}\PCOLDAGREE{\COMHWC{its}\HWC{with} consequential} physical phenomena necessary for the resolution of the paradox. 
For simplicity, we adopt the natural unit.

At the center of the paradox are three ``principles'' derived from well-tested theories of general relativity and quantum field theory: 1) The no-hair theorem \cite{Chrusciel:2012jk}\COMPCOLDAGREE{ that}\PCOLDAGREE{, which} suggests the mass $M$ as the only classical scale for a BH. For brevity, the charge and the angular momentum of BH are neglected. 2) The 1st law of BH thermodynamics \cite{Bardeen:1973gs}\COMPCOLDAGREE{ that}\PCOLDAGREE{, which} relates the observed mass change $\Delta M$ with the change of BH horizon area $\Delta A = 8\pi \kappa^{-1} \Delta M$ where 
$\kappa$ is the BH surface gravity. 3) 
The existence of Hawking radiation process \cite{Hawking:1974sw}\PCOLDAGREE{,} where 
a BH with a surface gravity $\kappa$ in the Unruh vacua (particle-less in-vacua \cite{Unruh:1976db}) emits thermal radiation of temperature $T_H=\kappa/ (2\pi)$.

Together, these principles paint the picture of a featureless BH that evaporates into a tremendous amount of radiation. This deduction alone is sound, as objects do radiate when burnt. The horizon even appears discretized, akin to Planck's oscillator, according to the 1st law\PCOLD{,} $\Delta M = T_H \Delta A/4 = T_H \Delta S_H$, where $S_H$ is the thermal entropy of BH deduced from the Hawking radiation. Each quantum with the entropy of one Hawking radiation particle occupies \COMPCOLDAGREE{a horizon area of four Planck area}\PCOLDAGREE{four units of Planck area
on the horizon}. \COMPCOLDAGREE{By i}\PCOLDAGREE{\HWC{By i}\COMHWC{I}}ntegrating \PCOLDAGREE{the first law}, we arrive at \PCOLDAGREE{the} Bekenstein-Hawking bound\PCOLD{,} $S_H \leq A /4$ \cite{Bekenstein:1973ur}, a.k.a. Bekenstein's law if the bound is saturated.
However, unlike ordinary thermal radiations whose entropy originates from the incomplete knowledge about the environment, Hawking radiation remains thermal regardless of the initial condition. 
Therefore the emitted radiation must be maximally entangled with 
whatever is inside BH.  Unless the horizon is somehow leaky, the entanglement entropy piles up and leads to a Planck-size BH \COMPCOLDAGREE{with}\PCOLDAGREE{that retains} a similar amount of entropy as the original one \cite{Hwang:2016otg}. This conclusion is in direct conflict with Bekenstein's law, and \PCOLDAGREE{the }
storage of the excess entanglement entropy \PCOLDAGREE{in a form }
other than \COMPCOLDAGREE{those}\PCOLDAGREE{the} horizon quanta, is necessary.

Notice that for observers outside BH, this new ingredient must be gravitationally inert for it to slip through the classical analysis. \HWC{More precisely BH should evaporate at a rate well approximated by the Hawking process.} 
While this ``hidden sector''\COMHWCOLD{ (HS)} \COMPCOLDAGREE{could}\PCOLDAGREE{can} represent various mechanisms, including the non-unitary process 
\cite{Unruh:1995gn,Unruh:2017uaw
}, intra-Hawking-radiation entanglement 
\cite{Parikh:2004ih,Almheiri:2012rt,Chiang:2020lem,Pasterski:2020xvn}, and physical entities \COMPCOLDAGREE{carrying entropy away}\PCOLDAGREE{that carry away the entropy} \cite{Chen:2014jwq,Hotta:2015yla,Carney:2017jut}, 
we \COMHWC{\COMPCOLD{must} emphasize}\HWC{reiterate} that our interest lies in the \PCOLDAGREE{generic} mathematical constraints laid down by those three principles. We will first focus on how \COMHWCOLD{HS}\HWCOLD{the hidden sector} interacts with BH \COMPCOLD{for}\PCOLD{to escort} entropy transfer, and then \PCOLDAGREE{on} whether Hawking temperature could be the statistical temperature of BH. From now on, unless stated otherwise, ``entropy'' refers to the entanglement entropy between BH and the exterior, \PCOLD{\COMHWC{which}\HWC{and} is} 
thus positive-definite and additive intra-BH.

\begin{figure}
\begingroup
\sbox0{\includegraphics{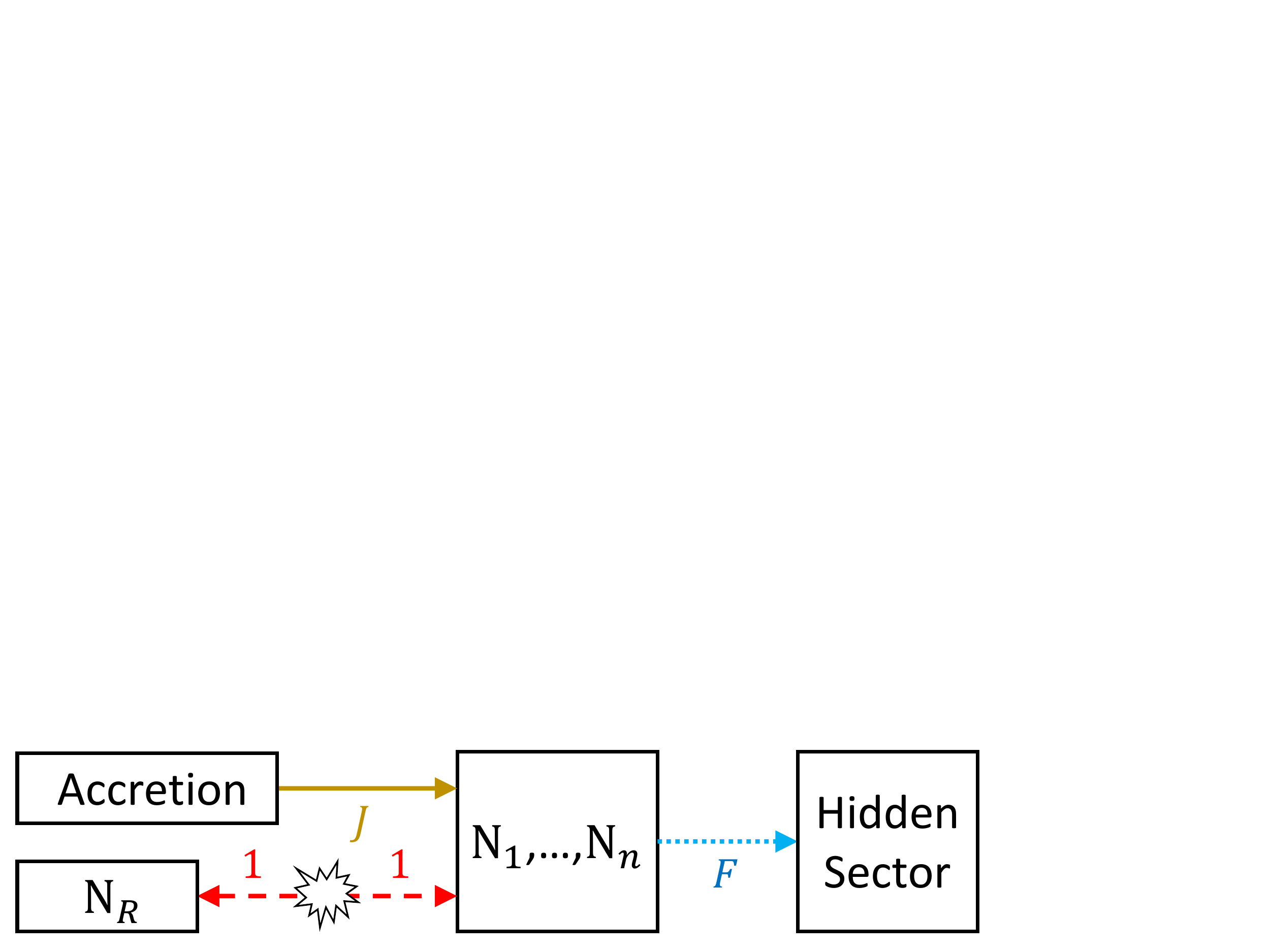}}%
\includegraphics[clip,trim={.01\wd0} {.01\ht0} {.23\wd0} {.77\ht0}, width = .48\textwidth]{interation.pdf}
\endgroup
\caption{\label{fig:horizon2}
Diagrammatic representation of the effective model. On the left are two observables: Hawking radiation $R$ from BH and accreted matter into BH. At the center is BH comprising $n$ different species sourced by the external \COMHWCOLD{current}\HWCOLD{source} $J$ (solid arrow) and the Hawking evaporation process (the bang and dashed arrows) that defines the unit of time as one increment of $N_R$ (Hawking radiation) per tick. The hidden sector on the right, sourced only by the hidden \COMHWCOLD{current}\HWCOLD{sink} $F$ (dotted arrows), is necessary for BH to lose entropy. Otherwise, the BH entropy, related to the sum of all species, accumulates, thus violating Bekenstein-Hawking bound. Notice that the inter-BH-species interactions are suppressed in this diagram.
}
\end{figure}

\paragraph*{Setup.}
A question 
arises when analyzing \COMHWCOLD{HS}\HWCOLD{the hidden sector}: what \PCOLD{entity} \HWC{does it} interact\COMHWC{s} with\COMHWC{ it}? \COMPCOLDAGREE{From}\PCOLDAGREE{Dictated by} the three principles, the only choice would be the horizon area quant\COMHWC{a}\HWC{um}\COMPCOLDAGREE{ that serve}\PCOLDAGREE{. We therefore invoke it} as the dynamical variable\COMPCOLDAGREE{s} of our effective model. 
With that in mind, we introduce a generic $(n+3)$-species macroscopic model\PCOLD{,} as depicted in \COMPCOLDAGREE{figure}\PCOLDAGREE{Fig.}~\ref{fig:horizon2}, \COMPCOLDAGREE{with}\PCOLDAGREE{that consists of} four major components: the Hawking radiation $R$, the external \COMHWCOLD{current}\HWCOLD{source} $J$, the horizon quanta\COMHWC{ \PCOLD{$N$} of BH}, and the hidden sector\COMHWCOLD{ HS}.
The horizon quanta are further categorized into $n$ species \HWC{of indistinguishable quanta}, labeled by \PCOLDAGREE{the} subscript $i=1\cdots n$\COMHWC{, \COMPCOLD{each}\PCOLD{where the quanta within the same species are} indistinguishable}\COMPCOLD{ from quanta of the same species}.
\COMPCOLD{(There is a lack of explanation about why you need to make such a category, and based on what criterion that you categorize them. What physical differences are these $n$ species?)}\COMHWC{There is no ``physical'' difference at this stage, only mathematical ones. The reason for multiple horizon quanta species is to encompass more possible solutions to ILP. With just 1 species (traditional 4 $L_P^2$ area quanta), there is only one valid solution, in which excess quanta are immediately transferred to HS (e.g. \cite{Hayden:2007cs}). To better emphasize this idea I move the following sentences from the next paragraph to here and added few more sentences.}
\HWCOLD{These\COMHWC{ abstract} species\COMHWC{ are the salient features of our model.} \HWC{enable our model to encompass scenarios other than those \cite{Hayden:2007cs,Gibbons:1976ue} strictly following the Bekenstein's law.} However, we must stress that \HWC{at this stage} they are just \COMHWC{mathematical}\HWC{abstract} constructs that characterize the model, \COMPCOLDAGREE{similar}\PCOLDAGREE{\COMHWC{which is }analogous} to the effective degrees of freedom in the mean-field theory.}
For simplicity, the \COMPCOLD{amount of}\HWC{amounts of} quanta and\COMHWC{ \PCOLD{the}} entropy 
\COMPCOLDAGREE{belonging}\PCOLDAGREE{that belong} to a species \COMPCOLD{$X$}\COMHWC{\PCOLD{$i$}}\HWC{$X$ ($R$ or $i$)}, both non-negative, are denoted as $N_{\COMPCOLD{X}\COMHWC{\PCOLD{i}}\HWC{X}}$ and $S_{\COMPCOLD{X}\COMHWC{\PCOLD{i}}\HWC{X}}$
.\COMPCOLD{(You have already introduced the notation $i$ for a species. So why do you now use capital $X$?) }\COMHWC{(We cannot use $i$ since the notation of $N$ and $S$ applies to Hawking radiation $R$ as well. Also we cannot rename $R$ as $0$ or any number that fits into $i$ since small Roman letters serve as the dummy indices for summing over the horizon quanta. Counting from $0$ or $1$ would definitely confuses reader more.)} \COMPCOLDAGREE{Furthermore, the}\COMHWC{\PCOLDAGREE{We further introduce a} clock\PCOLDAGREE{, which} synchronizes with the Hawking radiation $R$ by setting the emission rate to $1$. 
For brevity, }\HWC{In particular, $N_R$ serves as our clock. We further repurpose $\Delta$ as}\COMHWC{ from now on $\Delta$ denotes} the \COMPCOLDAGREE{change rate of the follow-up}\HWC{succeeding quantity's }\PCOLDAGREE{rate of change\COMHWC{ of the referred}}\COMHWC{ quantity}, e.g., $\Delta N_R = 1$\COMHWC{ according to the definition}.
\COMHWCOLD{These abstract species are the salient features of our model. However, we must stress that they are just mathematical constructs that characterize the model, \COMPCOLDAGREE{similar}\PCOLDAGREE{which is analogous} to the effective degrees of freedom in the mean-field theory. \COMHWC{(moved up)}} \COMHWCOLD{Specifically, the generating function\PCOLD{al} of the model depends only on $N$. \COMHWC{(moved to the next paragraph)}}

The only \COMHWC{physicality}\HWC{ad hoc} condition \COMHWC{we impose }\PCOLD{in the present work} is 
\COMPCOLDAGREE{the independence of}\PCOLDAGREE{that} the interaction strength \COMPCOLDAGREE{from}\PCOLDAGREE{is independent of} the entropy of each species, leaving the inclusion of entropic gravity \cite{Verlinde:2010hp} as future work. \HWCOLD{Specifically, the generating function\COMHWC{\PCOLD{al} (Since we only consider finite number of species so it is a function, not a functional.)} of the model depends only on $N_\HWC{i}$.} This assumption stems from the Weinberg-Witten theorem \cite{Weinberg:1980kq}, which suggests the non-compositeness of gravity that supposedly governs 
the \HWC{BH} microscopic degrees of \PCOLDAGREE{freedom\COMHWC{ of the}}\COMHWC{ BH}. 
As a result, any \COMPCOLDAGREE{entropy }transfer \PCOLDAGREE{of entropy} \HWC{originated} from a species $X$\COMPCOLD{  (to ...?)}\COMHWC{(The destination is irrelevant actually.)} can be associated with \COMPCOLDAGREE{a quanta transfer}\PCOLDAGREE{that of quanta,} 
as $N_X \Delta S_X = S_X \Delta N_X$. To ensure the \COMHWC{apparent }\COMPCOLD{(?)}\COMHWC{(apparent in the sense that the hidden sector may be representing unitarity violation. But to avoid confusion we can just remove the word)} unitarity, the total amount of quanta, including those transferred to \COMHWCOLD{HS}\HWCOLD{the hidden sector}, must be conserved. Consequently, the average entropy per quantum of BH \HWC{forever} \COMPCOLDAGREE{sits}\PCOLDAGREE{lies} between the\COMHWC{ initial} entropy per quantum of each \HWC{horizon quanta} species \HWC{at an instance}, \PCOLD{\COMHWC{i.e., }the}\COMHWC{ entropy per} Hawking radiation, and \COMPCOLD{the ratio between entropy and quanta}\COMHWC{\PCOLD{that of} }the external \COMHWCOLD{current}\HWCOLD{source}\COMHWC{ carries}. Surprisingly, even if the BH entropy in principle should be related to the quantum nature of BH and thus lies beyond our grasp, it is\COMHWC{ \PCOLD{actually}} bounded by a quantity proportional to the total amount of quanta $S^*$, which we utilize as the tracer. This affirms that \cite{Page:1993wv} for BH, the ``number of degrees \PCOLDAGREE{of freedom}'' 
still bounds its entropy.
However,\COMHWC{ \PCOLD{there is}} an additional assumption\COMHWC{ \PCOLD{that}} hides within the argument above. \COMPCOLDAGREE{In principle}\PCOLDAGREE{A priori}, there is no constraint on the content of the external \COMHWCOLD{current}\HWCOLD{source}. \COMPCOLD{The postulate}\PCOLD{\COMHWC{But in reality the}\HWC{The implicit} assumption} that the ratio is bounded, i.e., the holographic principle, provides \COMPCOLDAGREE{the}\PCOLDAGREE{a} bridge between the entanglement entropy and the thermal entropy of the 1st law. Notice that the bound \COMHWC{automatically exists}\HWC{emerges naturally} if \COMPCOLDAGREE{quanta transfers happen}\PCOLDAGREE{the transfer of quanta happens} discretely. 

\COMPCOLDAGREE{Now, let us}\PCOLDAGREE{Let us now} delve into the functionality of individual interactions in the system. First, a controllable external \COMHWCOLD{current}\HWCOLD{source} $J \equiv 1 + \Delta A /4$ represents the change to $M$ other than that due to the Hawking radiation, e.g., accretion. We only consider cases where entropy in the \HWCOLD{external} \COMHWCOLD{current}\HWCOLD{source} is homogeneous. To wit, it feeds into one particular species of quanta, though more complicated models are also possible. The second \COMHWCOLD{current}\HWCOLD{source} defines the effect of the Hawking radiation on BH, contributing $-T_H$ and $1$ \COMPCOLDAGREE{respectively }to $M$ and \COMPCOLDAGREE{total amount of quanta }$S^*$ \PCOLDAGREE{respectively}. This is \COMPCOLDAGREE{the}\PCOLDAGREE{a} manifestation of our obsession with energy conservation and unitarity. \COMPCOLDAGREE{In principle}\PCOLDAGREE{Na\" \i vely} these two \COMHWCOLD{current}\HWCOLD{source}s are the only observables of BH. But as argued before, there must exist a covert channel between BH and \COMHWCOLD{HS}\HWCOLD{the hidden sector} for the excess entropy to sink. First introduced in \cite{Hotta:2017yzk}
, this ``hidden \COMHWCOLD{current}\HWCOLD{sink}'' \COMHWCOLD{transfers}\HWCOLD{drains} a total 
amount of $F$ 
quanta \HWCOLD{from BH} to \COMHWCOLD{HS}\HWCOLD{the hidden sector}\COMHWC{, and would be the first target}\COMPCOLD{ (?)}. We will \HWC{focus on it and neglect the}\COMHWC{ skip over} intra-horizon-quanta interactions \HWC{for the moment} despite their necessity for the complete determination of the system evolution. \COMPCOLD{(?)}

\paragraph*{Quanta counting and \HWC{the} Bekenstein-Hawking bound}
To describe the resolution to \COMHWCOLD{ILP}\HWCOLD{the information loss paradox}, one must keep track of the horizon area and the total amount of quanta $S^* \HWC{\:= \sum_i N_i}$.
First, the area $A
$ must be 
a function of $N_i$ \COMHWC{that records the contribution to the horizon area by each species}\HWC{by construction}. 
In conjunction with the \COMHWC{massless}\HWC{inert}\COMHWC{ (inert is more consistent with previous notation)} condition of \COMHWCOLD{HS}\HWCOLD{the hidden sector}, the definition above leads to the evolution equation
\begin{align}
J - 1 = \Delta A/4 = 
\sum\nolimits_i \partial_{N_i} A\,
\Delta N_i /4  \,.  \label{eq:P}
\end{align}

Regarding\COMHWC{ the total amount of quanta} $S^*$, it is conserved, as argued before. Only the external \COMHWCOLD{current}\HWCOLD{source} $J$, the addition of a single quantum per tick due to the Hawking\COMHWC{ radiation} process, and the hidden \COMHWCOLD{current}\HWCOLD{sink} $F$ are capable of modifying $S^*$, resulting in $\Delta S^* \COMHWC{= \sum_i \Delta N_i} = \HWC{c_J^{-1}}J + 1 -F$\HWC{, where $c_J$ is the area per quantum of the external source}. Bekenstein-Hawking bound $S^* \propto S_{EE} \leq S_H \leq A/4
$ with a finite proportionality then can be expressed as the boundedness of $S^* / A
$, with
\begin{align}
\frac{d 
( S^*/A
) }{d \ln A
} = 
\frac{\HWC{c_J^{-1}} J-F+1}{4\left(J-1\right)} -\frac{S^*}{A
} \underset{J\to 0}{=} 
\frac{F-1}{4}-\frac{S^*}{A
}  \,.  \label{eq:SA}
\end{align}
We now realize that the introduction of the hidden \COMHWCOLD{current}\HWCOLD{sink} in \cite{Hotta:2017yzk} is not a coincidence, but rather out of necessity from the evolution of an entropy-density-bounding quantity. Without \COMHWC{the hidden \COMHWCOLD{current}\HWCOLD{sink}}\HWC{it}, $S^*/A
$ \PCOLDAGREE{would} diverge\COMPCOLDAGREE{s} as $A
$ approaches nullity, leading to \COMHWCOLD{ILP}\HWCOLD{the information loss paradox}. \COMPCOLD{(But a priori, Eq.(2) does not guarantee the cure of the divergence oroblem ad A goes to zero.!)}\COMHWC{(Do you mean $\delta A / A$ diverges as $A\to 0$? That is in quantum gravity regime and is out of scope.)}

Several conclusions can be drawn from the equation. First, the original Bekenstein-Hawking bound \cite{Bekenstein:1973ur,Page:1993wv} $S^* \propto A$ is satisfied only if $F \geq 1 + 4S^*/A
$ at a specific entropy density $S^*/A
\geq 1/4$. The factor $1+ 4S^*/A
$ coincides with simpler models \cite{Page:1993wv,Hwang:2016otg,Hotta:2017yzk} and represents both the excess entropy directly due to the Hawking radiation and indirectly due to the associated area reduction. 
Unfortunately, this 
constraint 
corresponds to a separatrix, which does not forbid BHs \PCOLDAGREE{that are} already on the other side from further deviating away, and most extremely, from forming remnants: \HWC{stable, nearly fully evaporated} BHs with a huge amount of \HWC{residue} entropy\COMHWC{ but with a vanishing horizon area}. Notice that remnants in our analysis are classical \cite{Page:1993wv,Chen:2014jwq} rather than of quantum gravity origin \cite{Adler:2001vs}. Those Planck-size remnants form only if quantum gravity corrections are introduced, \COMPCOLD{and are}\PCOLD{which is} beyond the semi-classical analysis here.

\paragraph*{Dimensional analysis and \HWC{the} extended Bekenstein's law}
\hspace{0.1pt}
\COMHWC{To better understand the interactions $\Delta N_i$
, we}\HWC{Let us switch gears, and} perform the dimensional analysis \HWC{on the previously dropped intra-horizon-quanta interactions (represented by $\Delta N_i$).} \COMHWC{and}\HWC{We} categorize \COMHWC{them}\HWC{the interactions} into marginal \COMHWC{interactions}\HWC{ones}\PCOLDAGREE{,} whose effects on the area $\partial_{N_i} A
\Delta N_i$ are roughly independent of $A
$\PCOLDAGREE{,} and (ir)relevant \COMHWC{interactions}\HWC{ones} 
that are (suppressed)enhanced
. For the moment, we \COMHWC{neglect}\HWC{turn off} the external \COMHWCOLD{current}\HWCOLD{source} $J$. Since the only other scale of the system is the Planck area, irrelevant interactions matter only in the quantum gravity regime \cite{Adler:2001vs,Chen:2014jwq,Hotta:2015yla}. For \PCOLDAGREE{the} relevant ones, the colossal difference between the Planck area and $A$ 
implies that some parameters must reach equilibrium and decouple from the rest rapidly.

A typical choice of parameters is the relative contribution of each species to the horizon area $z_i$, defined as
\begin{align}
\Delta z_i \equiv \left( \partial_{N_i} A
\Delta N_i - z_i \Delta A
\right) /A
\label{eq:zi}  \,.
\end{align}
$z_i$ has the benefit of being dimensionless, but in general one may choose whatever suitable.
For example, to characterize the asymptotic behavior of $S^*$ at the end of BH evaporation, the right-hand side of eq.~\eqref{eq:SA} can be parametrized as $f \left( z_i \right) 
A^k \sim f \left( S^*/A
\right) 
A^k$, given the monotonicity of the bounding function. The boundary between complete evaporation ($\ln S^* \to -\infty$) and potential remnant formation ($\ln S^*$ remains finite) happens at 
$f = - O \left( S^*/A
\ln\left( S^* /A
\right)^{-p} \right)$ with $p >0$ if $k=0$ and even weaker if $k>0$. Thus we arrive at \COMPCOLDAGREE{the}\PCOLDAGREE{a} sufficient condition that the entropy $S$ vanishes if asymptotically the hidden \COMHWCOLD{current flows}\HWCOLD{sink drains} at a rate
\begin{align}
F \gtrsim 4S^*/A
\,.  \label{eq:remnant}
\end{align}
In the case of \cite{Hotta:2017yzk}, the interaction is of the form $4\alpha S^*/A
$ with $\alpha \geq 1$, and remnants never form.


\COMHWC{Let us delve deeper into the classification of a species $j$ \COMPCOLDAGREE{by}\PCOLDAGREE{with} \COMPCOLDAGREE{its }property $Q$ (either the number of quanta $N_j$ or the area contribution $A
z_j$). Following the routine laid down previously, t}\HWC{Clearly, our model is governed by the relation between $A$ and $N_i$. To illustrate this, let us apply the routine laid down previously to the properties of a species $j$ ($N_j$ or the area contribution $A z_j$). T}he evolution equation of \HWC{a property} $Q$ is quantified as $\Delta Q \equiv -\left( Q/A
\right)^{1+p} 
A^k$\COMHWC{.}\HWC{,}
\begin{figure}
\HWC{\includegraphics[width = .47\textwidth]{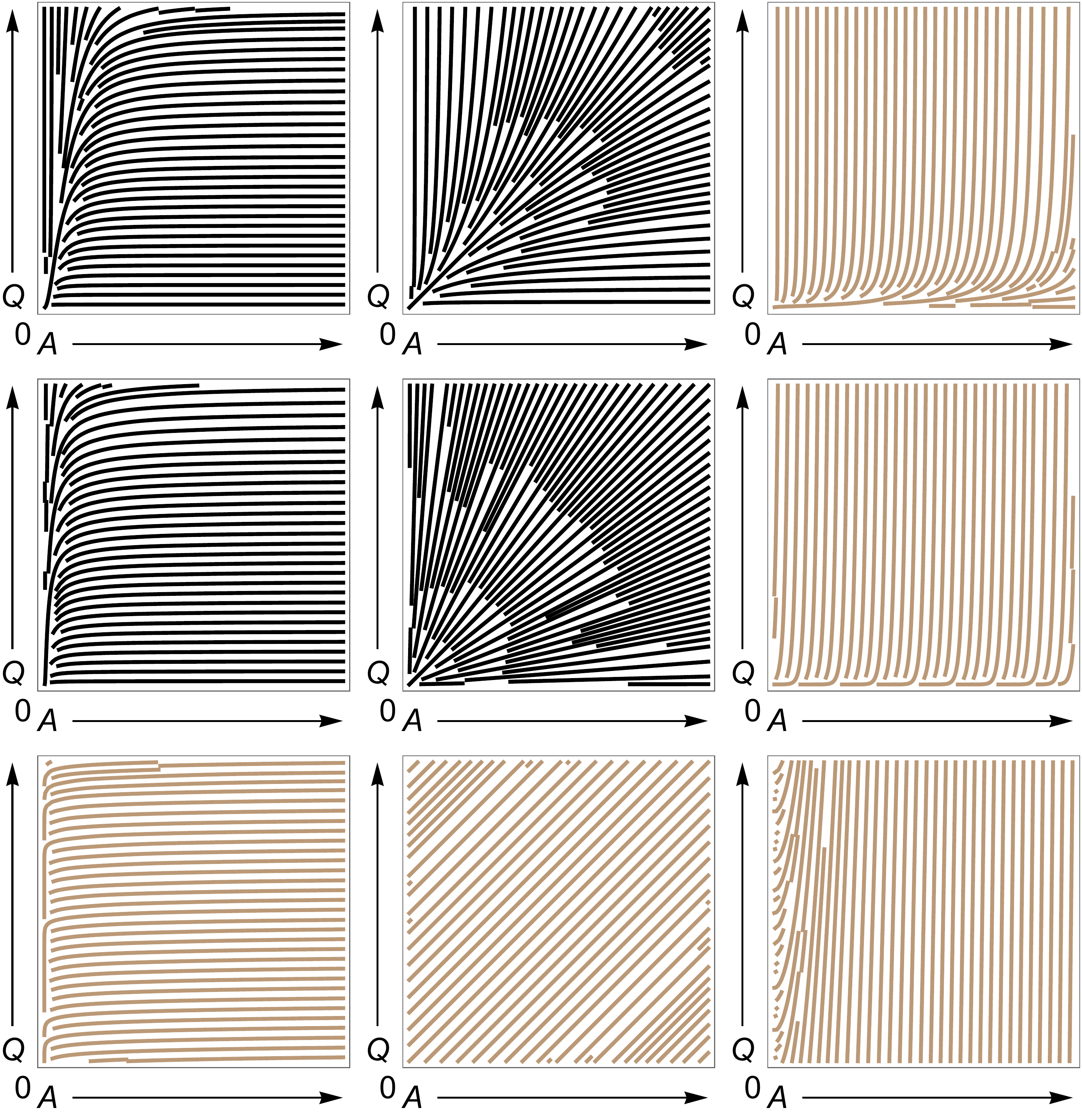}}
\caption{\label{fig:flow}
\COMHWC{Flow diagrams of $Q$ (vertical) versus $A
$ (horizontal)}\HWC{$Q-A$ flow diagrams} with $\Delta Q \equiv -\left( Q/A
\right)^{1+p} 
A^k$ and $\Delta A
= -\COMHWCOLD{1}\HWCOLD{4}$. \HWC{The origin is labelled with $0$.} Each diagram corresponds to a different set of parameters, with from bottom to top $p = -1,0,1$ and from left to right $k = -1,0,1$. Only \COMHWC{for the top right 4 cases}\HWC{cases marked in black} ($p \geq 0$ and $k \COMHWC{\geq}\HWC{\leq} 0$) \HWC{meet our criteria.}\COMHWC{$Q$ decays fast enough to prevent remnants from forming before BH reaching Planck scale, yet slow enough to avoid divergence. In the cases with $k > 0$ however, $Q$ decays inverse polynomially in $A
$ and equilibrates at $Q
\ll A$, thus rendered negligible.}
}
\end{figure}
\COMHWC{\COMHWC{(These are three different scenarios, not two.)} Most of the evolution trajectories 
either lead to remnant formation, \COMPCOLD{a constant}\PCOLD{where} $Q$ \PCOLD{remains constant} till $A
\sim 1$ (final burst), or \COMPCOLD{unavoidably diverging} $Q$ \PCOLD{inevitably diverges} toward $-\infty$, as demonstrated in Fig.~\ref{fig:flow}, except when either $k=0$, $p \geq 0$ (marginal) or $k>0$, $p \geq 0$ (inverse polynomial decay). 
For the second scenario}\HWC{and the trajectories, as demonstrated in Fig.~\ref{fig:flow}, behave wildly differently depending on $p$ and $k$. When $p<0$, $Q$ diverges toward $-\infty$, and if $k>0$}, $Q$ is quickly attracted toward a \COMHWC{nearly vanishing}\HWC{pure} function of $A$\HWC{, and is no longer an independent parameter. In addition, for $A z_j$, since the overall evolution $\Delta A =4$ is marginal, a species with $k<0$ must be accompanied by yet another species with $k<0$, such that the net area contribution remains marginal}. Thus we may treat $k \sim 0$ for\COMHWC{ the evolution of} $A z_j$, and categorize species according to $k$ of $N_j$ \HWC{alone}:\COMHWC{ $k>0$ 
($N_j$ approaches the attractor, rendering $z_j$ a pure function of $A
$ and thus not an independent parameter),} 
$k<0$ (final burst), and $k=0$ (marginal) where $\Delta N_j$ and $\Delta A
$ are effectively dimensionless, leading to $\partial A
/ \partial N_j = c_j$ where $c_j$ are constants, i.e., 
\begin{align}
A/4
= \sum\nolimits_i c_i N_i  \,.  \label{eq:quanta}
\end{align}

This ``extended'' Bekenstein's law allows us to apply the horizon quanta picture to \COMHWC{all}\HWC{marginal} species, each with a different area per quantum $c_i$. 
Furthermore, $c_i$ also indicates the existence of a 
minimal quanta area density $S^*/A
$, i.e., the inverse of the maximally possible area per quanta $1/ (4\max_i c_i )$. As \PCOLDAGREE{will be} demonstrated later, this lower bound is related to the dimensionality of the microscopic degrees of freedom through the statistical analysis.

\paragraph*{Generating function, density of states, and soft hairs}

By \COMPCOLDAGREE{definition}\PCOLDAGREE{construction}, all observables in our model are functions of $N_i$. Namely, the connected generating function of BH can be expressed as $\ln Z_B (N_i) \equiv \ln Z_B (\beta, N_i | \beta)$, where $\beta = T_H^{-1} 
$ and $N_i | \beta$ is $N_i$ under the constraint of $\beta = \sqrt{4\pi A(N_i)}$.
While the full expression of $\ln Z_B$ requires quantum gravity, we may qualitatively determine the part related to $\beta$ given our knowledge about the rest of the system. In fact, both the external \COMHWCOLD{current}\HWCOLD{source} and \COMHWCOLD{HS}\HWCOLD{the hidden sector} are trivial in terms of the energy content as the former is external and the latter is gravitationally inert\COMHWC{, i.e., non-energetic}. The Hawking radiation alone suggests $\beta$ as the conjugate variable of the microscopic total energy $E$, with an effective action $\Gamma_B (E) = \max_{\beta} \{\beta E + ln Z_B(\beta)\}$.
The origin of conjugation is beyond us. We consider it as the thermodynamic conjugate \cite{Hotta:2017yzk}, but other explanations are also plausible \cite{Cotler:2016fpe,Gibbons:1976ue}. $Z_B$ then becomes the ensemble average $\int e^{\Gamma_B(E) -\beta E} dE$, with $\ln Z_B \sim -\beta^2/(16\pi)$ from $M \sim \langle E \rangle = - \partial_\beta \ln Z_B$. The Gaussian form of $Z_B$ suggests BH as an effective single-particle state \cite{Gibbons:1976ue,Chung:2018kqs}.\HWC{

}However, as an assumption of our model, and given its entropic nature, BH should be composite. The missing piece is that the Hawking process not only modifies $M$ but also the amount of quanta $N$, rendering $Z_B$ unattainable again. The only exception happens when \COMHWC{$N$ equilibrates, e.g., when BH is in Hawking vacua. The single-quantum density of states $g$\COMPCOLDAGREE{ then} can \PCOLDAGREE{then} be extracted from}\HWC{a particular energy measurement becomes independent of $N$. For a marginal species $i$ satisfying eq.~\eqref{eq:quanta}, $\partial_{N_i} M \propto M^{-1}\partial_{N_i} A$ depends only on $\beta$, indicating that} the average of the single-quantum energy $\omega_\HWC{i}$ \HWC{can be utilized to extract the single-quantum density of state $g_i$} through
\begin{align}
\left\langle \omega \right\rangle_\HWC{i} \HWC{\:=\frac{A z_i}{2 N_i \beta\:}=\frac{2c_i}{\beta}\,}\equiv 
- \partial_\beta \ln \int_0^\infty g_\HWC{i} (\omega) e^{-\beta \omega} d\omega
\COMHWC{\equiv \frac{- \partial_\beta \ln Z_B}{N (\beta)}  + \mu
\,,}\HWC{\,.}  \label{eq:densityofstate}
\end{align}
\COMHWC{where $\mu$ is the chemical potential. }
We will restraint ourselves from solving \HWC{the associated chemical potential} $\mu_\HWC{i}$ as all we care about is its existence inherited from the\COMPCOLDAGREE{ quanta} conservation \PCOLDAGREE{of quanta} set up before, and thus the capability of expressing most forms of $N_\HWC{i} (\beta)$.\COMHWC{

} \HWC{By reversing the relation between $c_i$ and $g_i$ we arrive at $g_i = \omega^{2c_i - 1}$. This is typical for relativistic particles inside a $2c_i$-dimensional configuration space\COMHWCNEW{ \cite{Almheiri:2019hni}}.\COMHWC{ (moved from next section.)}} For\COMHWC{ example,} BHs following the original Bekenstein's law $S^*=A/4
$\COMHWC{would have the relation reduced to
\begin{align}
\COMHWC{\langle\omega\rangle = 4M/A
= 2/\beta \,,\quad
g(\omega) = \mathcal{L}_{\omega,\beta}^{-1} \left[ \beta^{-2} \right] = \omega \,,  \label{eq:DOFordinaryBH}}
\end{align}
where $\mathcal{L}_{\omega,\beta}^{-1}$ is the inverse Laplace transform from $\beta$ to $\omega$. A linear density of states (typical for a 2\COMPCOLDAGREE{$-D$}\PCOLDAGREE{D} relativistic system) implies that these} \HWC{the observation implies that the BH} quanta live on the horizon, a picture usually depicted as the salient feature of BH.
\COMHWC{Apparently $g(\omega)$ is fully determined by the relation between $\langle\omega\rangle$ and $\beta$, but what we considered as the average single-quantum energy depends on what is in equilibrium. 
Given a quanta counting $\tilde{N}(\beta)$ with $\Delta \tilde{N} = 0$ and the associated energy measurement $\tilde{M}(\beta)$, we have
\begin{align}
\COMHWC{g(\omega) \propto \mathcal{L}_{\omega,\beta}^{-1} \left[ \exp \left( \int_\beta^\infty \frac{\tilde{M}\left( \beta' \right)}{\tilde{N} \left(\beta' \right)} d\beta' \right) \right]  \,.  \label{eq:dos}}
\end{align}
For eq.~\eqref{eq:dos} to be physical, the argument of $\mathcal{L}_{\omega,\beta}^{-1}$ must be completely monotonic, i.e., $g(\omega) \geq 0$. $\tilde{N}$ thus grows at least as fast as $\beta \tilde{M}$ when $\beta \to \infty$. If $\tilde{N}$ is of the order $\beta \tilde{M}$ the asymptotic proportionality would be the dimension of the quanta configuration space, c.f. eq.~\eqref{eq:DOFordinaryBH}. Otherwise, the exponent vanishes, and by}\HWC{

For $k<0$ (final burst) species, it is not apparent how one would factorize the generating function. Nevertheless, we may still apply eq.~\eqref{eq:densityofstate}. By} the initial value theorem, there would be a pole at zero frequency inside the density of states. \HWC{However, we must emphasize that given the non-vanishing area contribution, this zero-frequency pole is just an illusion, and its emergence should be considered as an appeal against an arbitrary factorization. Interestingly, soft hairs \cite{Strominger:2013jfa,Hawking:2016msc,Hawking:2016sgy} are often considered a non-energetic entropy storage, capable of memorizing the evaporation history, as does the zero-frequency pole. Another resemblance between the two is that both of them are non-trivial around zero frequency \cite{Chiang:2020lem,Flanagan:2021ojq}. In fact, our analysis suggests the non-trivial time dependence as the key to the soft hair proposal.} Whether this structure \COMHWC{is related to}\HWC{has implication on} soft hairs requires further analysis.

\HWC{Given our blunt force of eq.~\eqref{eq:densityofstate} upon $k<0$ species, one wonders if we may simply consider the entire BH, with $\langle \omega \rangle =  A/(2S^*\beta)$. It surely can be done.}\HWCOLD{\COMHWC{(moved from next paragraph)} Since $S^*/A
$ does not increase once \COMPCOLD{reaching}\PCOLD{it reaches} the separatrix (
equilibrium), $A
/S^*$ should be a non-increasing function of $\beta$, \COMPCOLD{leading}\PCOLD{which, as a consequence, leads} to a system at most as energetic as the one with dimension $d \leq \max \{ A
/S^* \}/2 = 2\max_i c_i$. Furthermore, the constraint $S^*/A
\geq 1/4$ at the separatrix ensures \PCOLD{that} $d \leq 2$. The\COMHWC{se} addition\COMHWC{al species} \HWC{of species for the resolution of the information loss paradox} thus reduce $\langle\omega\rangle$, or increase $S^*$ at fixed $\beta$, as expected.}

\paragraph*{\COMHWC{Dimensionality and n}\HWC{N}on-commutativity of the configuration space}
\COMHWC{Let us be more specific about the choice of equilibrium. Na\" \i vely, one may consider Hawking vacua with $\tilde{M} = M$ and $\tilde{N} = S^*$, i.e., equilibrium between all species. In this case, the integrant inside eq.~\eqref{eq:dos} becomes $A/(2S^*) \, \beta^{-1}$, thus related to the dimensional analysis.}
\COMHWCOLD{Since $S^*/A
$ does not increase once \COMPCOLD{reaching}\PCOLD{it reaches} the separatrix (
equilibrium), $A
/S^*$ should be a non-increasing function of $\beta$, \COMPCOLD{leading}\PCOLD{which, as a consequence, leads} to a system at most as energetic as the one with dimension $d \leq \max \{ A
/S^* \}/2 = 2\max_i c_i$. Furthermore, the constraint $S^*/A
\geq 1/4$ at the separatrix ensures \PCOLD{that} $d \leq 2$. These additional species thus reduce $\langle\omega\rangle$, or increase $S^*$ at fixed $\beta$, as expected.}\COMHWC{(Moved to previous paragraph)}\COMHWC{\PCOLD{

}Another way \COMPCOLD{(of what?)} is to dissect BH into individual species $i$ equilibrated by the external \COMHWCOLD{current}\HWCOLD{source}, substituting $S^*$ with $N_i$ and $A$ with area contribution $A z_i$.\COMPCOLD{(This sentence is so bad that I cannot even understand.)} For species with $k=0$, eq.~\eqref{eq:quanta} indicates that $2c_i$ 
is the dimension of the configuration space
\cite{Almheiri:2019hni}. The other possibility is the zero-frequency pole appeared if $N_i$ grows faster than $\beta \tilde M \sim A z_i$. By parametrizing $A z_i = N_i^{1-x}$ with $x>0$ and $\Delta (A z_i) = -z_i^{1+p}$ with $p \geq 0$, we have $\Delta N_i 
= - \left(N_i/A \right)^{1+p} N_i^{x(-p)} \equiv - \left(N_i/A \right)^{1+p} N_i^k$, i.e., $k \leq 0$. Only species that keep accumulating (final burst) can populate the pole. Conversely, $k>0$ would 
lead to $N_i$ growing slower than $A z_i$, a case deemed impossible by the positivity of $g$. This is as expected since species with $k>0$ are fully determined by $A$ at the attractor, thus unable to reach equilibrium on their own.}
\COMPCOLD{(Is the above paragraph necessary? I would prefer to offer only one scenario.)}
While the analysis above is tempting, it is dangerous since to accommodate the effect of the Hawking evaporation process, \COMPCOLDAGREE{at least a species has a spectrum}\PCOLDAGREE{there would be at least one species whose spectrum is} unbounded from below. To avoid the runaway \COMPCOLDAGREE{effect}\PCOLDAGREE{situation}, the net energy transfer from these species to the others must be finite. Effectively, we have two well-separated subsystems, one containing all negative\COMHWC{ly }\HWC{-}mass\COMHWC{ed} quanta with total mass $M_n$ and the other containing the rest. 
%
%
%
%
\COMPCOLDAGREE{Consequently}\PCOLDAGREE{As a consequence}, after a while, the energy transfer rate from the negative\COMHWC{ly }\HWC{-}mass\COMHWC{ed} subsystem to the rest (excluding Hawking radiation), denoted as $\rho_n$, has to be negative-definite. 
\COMHWC{\\\indent
}By parametrizing $A z_i = 4 c_i N_i^{1-x_\HWCOLD{i}}$ with $x_\HWCOLD{i}<1$ \HWC{\footnote{\HWC{Through some manipulations, $x>1$ maps to $p<0$, which is forbidden.}}} \HWCOLD{and $c_i \neq 0$},
\begin{align}
-\Delta M_n &=
-\sum\nolimits_i \left(4 c_i\right)^{-1} N_i^{x_\HWCOLD{i}} [\partial_{N_i} M_n] ( N_j ) [\Delta (A z_i)] ( z_j ) \nonumber  \\
&= \rho_n + T_H \leq T_H
= \left( 16\pi c_i N_i^{1-x} /z_i \right)^{-1/2}  \,,
\end{align}
can be transformed into an inner product between a vector independent of $N_i$ and another \PCOLD{one} independent of $z_i$. 
One then applies the separation of variables and obtains the relation between the upper and lower bounds of two vectors along variables $z_i$ as
\begin{align}
1 / \max \left( \sqrt{\pi c_i / z_i} c_i^{-1} \Delta (A z_i) \right) &\geq \sqrt{N_i^{1+x_i}} \partial_{N_i} M_n  \,,  \nonumber\\
1 / \min \left( \sqrt{\pi c_i / z_i} c_i^{-1} \Delta (A z_i) \right) &\leq \sqrt{N_i^{1+x_i}} \partial_{N_i} M_n  \,.
\end{align}

The \COMHWCOLD{upper}\HWCOLD{lower} bound indicates $\partial M_n/ \partial N_i \COMHWCOLD{\leq}\HWCOLD{\:\geq\:} 0$ for a system satisfying eq.~\eqref{eq:remnant}, i.e., without remnants, while the \COMHWCOLD{lower}\HWCOLD{upper} bound suggests $\partial M_n/ \partial N_i \COMHWCOLD{\geq}\HWCOLD{\:\leq\:} 0$ if $c_i^{-1} \Delta (A z_i) > 0$ at $z_i=0$. Notice that $\Delta (A z_i) = 0$ at $z_i=0$ is a trivial fixed point \HWC{where $N_i$ vanishes indefinitely}\COMHWC{where the \PCOLD{$i$-th} species \COMPCOLD{$i$ }does not exist unless the area contribution $c_i$ vanishes}. \COMHWC{If $c_i$ does vanish, \PCOLD{then} the \PCOLD{$i$-th} species \PCOLD{would} decouple\COMPCOLD{s} from the exterior observers completely, and therefore by Occ\COMPCOLD{u}\PCOLD{a}m's razor, we do not consider it physical.}\HWC{For species with $c_i=0$, they decouple from the exterior observers and become spurious by Occam's razor.} $\partial M_n/ \partial N_i$ thus is positive definite unconditionally. Together\COMPCOLD{ we have}\PCOLD{, we assert that} $\partial M_n/ \partial N_i=0$, i.e., $M_n = 0$\COMHWC{, if} \HWC{for BHs that would not evolve into }remnants\COMHWC{ with large amounts of residue entropy never form}.

$M_n=0$ forbids the negative\COMPCOLDAGREE{ly }\PCOLDAGREE{-}mass subsystem from sourcing $A$, \PCOLDAGREE{and is} thus held at the trivial fixed point $z=0$. We \PCOLDAGREE{therefore} have no choice but to consider $x=1$ for \PCOLDAGREE{the} negative\COMPCOLDAGREE{ly }\PCOLDAGREE{-}mass species, i.e., a constant area contribution $A_n$ of the order of \PCOLDAGREE{the} Planck area.
Equipped with that knowledge, we may derive the density of states for the other half of BH
\begin{align}
g \left( \omega \right) \propto \omega ^{\COMHWC{c}\HWC{d/2}} j_{\COMHWC{c}\HWC{d/2\:}-1}(\sqrt{-A_n}\omega ) 
\,,
\end{align}
where $j$ is the spherical Bessel of the 1st kind 
and \COMHWC{$4c$}\HWC{$d$} is\HWC{, as defined previously,} the \COMHWC{area per quantum}\HWC{dimension of the configuration space} for the \COMPCOLDAGREE{positively massed}\PCOLDAGREE{positive-mass} subsystem.
This is typical for fields 
in a non-commutative space \cite{Connes:1996gi,Doplicher:1994tu,Batista:2002rq,Chiang:2015pmz}, 
\COMPCOLD{hinting at}\PCOLD{which hints at the existence of} a microscopically discrete horizon 
\cite{Hashimoto:1999ut,Ryu:2006bv}, despite the semi-classical nature of our analysis.

\paragraph*{Remark}
\COMHWC{In summary, we \PCOLD{have} construct\PCOLD{ed} a generic macroscopic model for \PCOLD{the} unitary \COMPCOLD{BH }evolution \PCOLD{of BH} based solely on the horizon quanta picture of BH thermodynamics. We then conduct\PCOLD{ed} various qualitative analyses to narrow the parameter space. Finally, we relate\PCOLD{d} these macroscopic parameters to the properties\COMPCOLDAGREE{ of the corresponding microscopic model}, including zero-frequency poles and UV cutoff\PCOLDAGREE{,  of the corresponding microscopic model}. This work represents the best effort we can accomplish without laying out the model in detail.}\COMPCOLD{ (I don’t like this sentence, but cannot come up with a replacement so far. I would much prefer to have one or several sentences that recast the key concepts of our theory. Please give it a try.)} \HWC{The concept and the framework laid down in this work is extremely general. One may even apply the ``hidden sector'' concept to the analog black hole model \cite{Unruh:1980cg,Steinhauer:2014dra,Chen:2015bcg}, and consider the interaction between the microscopic degrees of freedom in the analog horizon and the analog Hawking radiation as a probe to either the horizon, the radiation, or the expected Hawking radiation partner after detecting the other two. Furthermore, by specializing in a particular mode, more information can be extracted. For example, by applying the analysis of the last paragraph to the model in \cite{Hotta:2017yzk}, one realized that it only makes sense if the energy spectrum is bounded from both below and above, thus incapable of describing a perfect thermal radiation\COMHWCNEW{, i.e., Hawking radiation}. This is also what prompted us to consider a more generic model in the first place. \HWCNEW{In contrast, the additional species in \cite{Almheiri:2019hni,Penington:2019kki} occupies a fixed area, and is compatible with our analysis.} One may also consider our model from the information-theoretic point of view \cite{Hayden:2007cs}, by treating quanta as qubits. We have to, unfortunately, leave it as a future work.}

DY is supported by the National Research Foundation of Korea (Grant no.
2021R1C1C1008622, 2021R1A4A5031460). HW and KY are supported by Ministry of Science and Technology of Taiwan\COMHWC{, Center for Theoretical Physics (CTP)} and the Leung Center for Cosmology and Particle Astrophysics (LeCosPA) of National Taiwan University. HW would also like to appreciate the discussions with \HWC{Masahiro Hotta,} Feng-Li Lin, Wei-Hsiang Shao, and Naoki Watamura.


\bibliography{bibliography}



\end{document}